# CAUSAL STRUCTURES IN LINEAR SPACES

**Victor R. Krym**
*Dept. of Geometry, V.A.Steklov Mathematical Institute, St.-Petersburg Division,
Fontanka 27, St.-Petersburg 191011, Russia.
E-mail: vkrym@pdmi.ras.ru*

*Abstract.*
Linear kinematic spaces (i.e. linear topological spaces with partial ordering) are studied. They are defined by a set of 8 axioms implying that topology, linear structure and ordering are compatible with each other. Proved that the weakest topology of a linear kinematics is determined by the open cone of future of any its point. This topology allows decomposition of a linear kinematics in a Cartesian product of an Euclidean, antidiscrete and discrete spaces. Proved that the structure of linear kinematics is preserved by factorization along the antidiscrete component of this decomposition.
Linear topological spaces with pseudometric which satisfies time inequality instead of the triangle inequality are studied (3 axioms). Pseudometric (which is determined by a pseudonorm) is shown to define a topology on a linear space, it being a continuous mapping in this topology. Proved that for a space with pseudometric to be a linear kinematics it is necessary and sufficient that mapping of multiplication by −1 be continuous. Pseudometric topology of a linear kinematics is identical with the weakest topology of a linear kinematics with the given cone of future. Minkovskii space of the special relativity theory proved to be a pseudometric linear kinematics.

# 1. Introduction

**1.1.** There are no doubts that the idea of causality is one of the main principles of physics. However the appropriate mathematical relation defined on some space typical for the relativity theory (e.g., Minkovskii linear space or a pseudo-Riemannian manifold) was studied only in 1960s by H. Busemann [3] and R.I. Pimenov [5]. After this time only a few scientists paid attention to this mainly geometrical subject. However both Busemann and Pimenov missed important features of possible systems of axioms developed for description of the causal relationship. They do not discuss non-Euclidean topologies and they assumed that pseudometric (which do not satisfy the triangle inequality) does not determine a topology on a given space. In this paper we begin our analysis of causal structures defined on spaces relevant for the relativity theory. We discuss spaces which simultaneously possess three structures: the structure of linear space $L$ over the field $\mathbb{R}$, topological structure (the system $\Delta$ of open sets) and partial ordering. In physics the latter is interpreted as causal structure. We are interested in properties which can be derived from the following system of axioms.

**Definition.** *Linear space $L$ with topology $\Delta$ and partial ordering $<$ is called linear kinematics, iff it satisfies the following three groups of axioms.*

**1.2.** First group of axioms.

**H$_1$.** *The mapping of addition $+: L \times L \to L$ is continuous.*

**H$_2$.** *For any $\alpha \in \mathbb{R}$ the mapping of multiplication by $\alpha$ is continuos.*

It can be easily deduced that for any $a \in L$ and for any open set $U$ its translation $a+U$ is open, the translation of a closed set is closed, homothety $a \mapsto \alpha a$ with non-zero coefficient $\alpha$ transfers open set onto open set and closed set onto closed set. The axiom H$_2$ is much weaker than the usual requirement that the mapping of multiplication *as a whole* (i.e. $\mathbb{R} \times L \to L$) be continuous. We shall not require this.

**1.3.** Second group of axioms.

**O$_1$.** *Partial ordering $<$ is transitive: from $a<b$ and $b<c$ follows that $a<c$.*

**O$_2$.** *Partial ordering $<$ is strict: $a<a$ is impossible.*

**1.4.** Third group of axioms.

**LK$_1$.** $\forall a, b \in L \quad a<b \implies \forall u \in L \quad a+u<b+u$.

**LK$_2$.** $\forall a, b \in L \quad a<b \implies \forall \alpha \in \mathbb{R}_+ \quad \alpha a < \alpha b$, *where* $\mathbb{R}_+ := (0, +\infty)$.

**LK$_3$.** $\forall a \in L \quad \forall U \in \Delta \quad a \in U \implies \exists b \in U \quad a<b$, *i.e. in any neighbourhood $U$ of any point $a$ there is at least one point which follows $a$ (next point).*

**LK$_4$.** *The past of the zero point $Q_0^- := \{u \in L \mid u<0\}$ is open.*

Axioms H$_1$, H$_2$, LK$_1$, LK$_2$ imply that the topology $\Delta$, linear structure and ordering ($<$) are compatible with each other. Axiom LK$_3$ implies that the relation of ordering is topologically non-empty. Axiom LK$_4$ implies that all points which are close enough to some point in the past of the zero point also belong to the past of the zero point.

We do not imply that $L$ has Euclidean topology as it was assumed by Pimenov in [5]. It creates two problems. 1. To find which topologies are compatible with the given set of axioms. 2. To determine which structures introduced in [5] are preserved by this set of axioms.

# 2. Corollaries

**2.1.** LK$_3$ implies that the topology in $L$ is not discrete. Otherwise there would be no next point in a one-point neighbourhood of a given point. LK$_4$ implies that the topology in $L$ is not antidiscrete.

**2.2.** $\forall a, b \in L \quad a<b \implies \forall \alpha \in \mathbb{R}_- \quad \alpha a > \alpha b$, where $\mathbb{R}_- = (-\infty, 0)$.

(We skip proofs in this part due to their simplicity).

**2.3.** $\forall a \in L \quad \forall U \in \Delta \quad a \in U \implies \exists c \in U \quad c<a$, i.e. in any neighbourhood $U$ of any point $a$ there is at least one preceding point.

**2.4.** For any point $a \in L$ its past $Q_a^- := \{u \in L \mid u < a\}$ and its future $Q_a^+ := \{u \in L \mid u > a\}$ are open sets.

**2.5.** $\forall a, b \in L \quad a < b \Rightarrow Q_a^+ \supset Q_b^+, Q_a^- \subset Q_b^-$. It follows from the transitivity of order.

**2.6.** For any $a \in L$ the sets $Q_a^+, Q_a^-$ are convex.

**2.7.** For any $a \in L$ the sets $Q_a^+, Q_a^-$ are cones with apex $a$. It follows from the easily verified fact that $Q_0^+$ is a cone with apex 0. $O_2$ implies that apexes do not belong to cones.

### 3. Possible topologies

**3.1.** In this part we shall answer the question which topologies are compatible with the given system of axioms for finite-dimensional linear kinematics. Since convex cones $Q_0^+$ and $Q_0^- = -Q_0^+$ are open and the property of a set to be open is preserved by translations, one can easily find the weakest topology in the finite-dimensional linear space $L$, which is compatible with these two facts. It is obvious that the weakest topology in $L$ is closely related with the type of the cone $Q_0^+$. Let us begin with the following note.

**3.2.** If aff $Q_0^+ \neq L$, then the space $L$ can be decomposed to a Cartesian product of the subspaces $L' = $ aff $Q_0^+$ and $M = L/L'$. The ordering and the weakest topology in $L'$ are defined by the cone $Q_0^+$, the $L'$ and all its translates are open, hence the topology in $M$ is discrete. Therefore it is sufficient to find an answer to the question formulated in 3.1 for the case where aff $Q_0^+ = L$.

Let us first consider the spaces of small dimensions.

**3.3.** Let $L = \Re^1$. Let us simulate $\Re^1$ by the points of the real axis $\mathbb{R}^1$ and let $Q_0^+$ be an open ray in the standard (Euclidean) topology of $\mathbb{R}^1$. Then the ray $Q_0^-$ is also open. The intersections of their translates are also open, i.e. all segments $a < t < b$ are open. Therefore the weakest topology in $L$ is Euclidean. It is compatible with the axioms of part 1.

**3.4.** Let $L = \Re^2$. Simulate $\Re^2$ by the points of a Cartesian plane $\mathbb{R}^2$ (preserving linear structure). We shall consider 5 cases:

1. $Q_0^+$ is an open in $\mathbb{R}^2$ sharp (i.e. it does not contain straight lines) convex sector with angle $< \pi$.

2. $Q_0^+$ is an open in $\mathbb{R}^2$ convex sector containing straight lines, i.e. an open half-plane.

3. $Q_0^+$ is a sector like (2) with addition of one of its edge rays.

4. $Q_0^+$ is a sector like (1) with addition of one of its edge rays.

5. $Q_0^+$ is a sector like (1) with addition of both of its edge rays. In the case (2) the addition of both rays is impossible due to axiom $O_2$.

**3.4.1.** In the case (1) the weakest topology is Euclidean. If we introduce coordinates $(x, t)$ on the Cartesian plane $\mathbb{R}^2$ directing $t$ axis along the bisectrix of the cone $Q_0^+$, we shall see that it is a model of two-dimensional space-time of the special relativity theory. All axioms of part 1 are fulfilled.

**3.4.2.** In the case (2) let us introduce Cartesian coordinates $(x, t)$ on the plane $\mathbb{R}^2$ directing $x$ axis along the edge of the half-plane $Q_0^+$. The weakest topology can be defined if we choose all strips $a < t < b$ as the base of topology. Topology is different from Euclidean, $L = \mathbb{R}^1 \times \mathbb{R}^1_a$, where $\mathbb{R}^1_a$ is the real axis with *antidiscrete* topology. It is a model of two-dimensional Halilean space-time. All axioms of part 1 are fulfilled.

**3.4.3.** In the case (3) let us introduce coordinates $(x_1, x_2)$ on the plane $\mathbb{R}^2$ directing $x$ axis along the edge of the half-plane $Q_0^+$, its direction being the same with the direction of the ray which is included in $Q_0^+ = Q_{(0,0)}^+$, Fig.1.

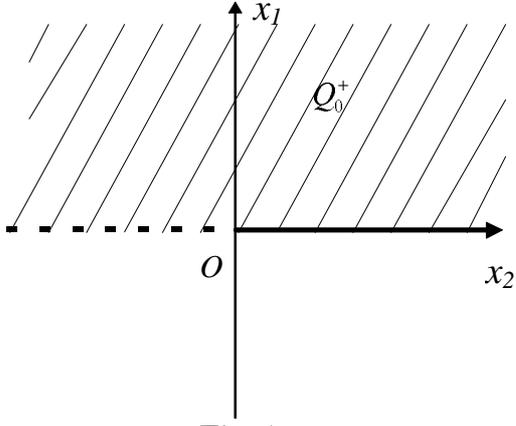
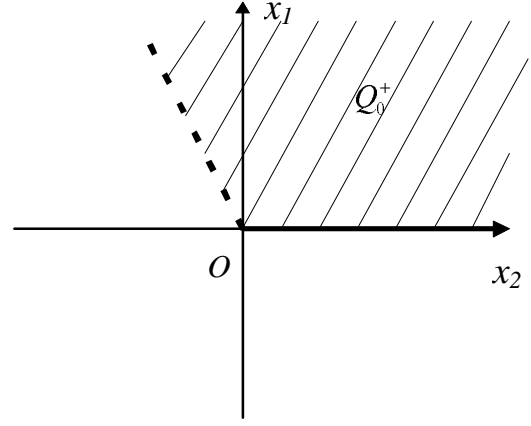

Fig. 1.                Fig. 2.

Ordering which is introduced by this cone corresponds to a strict lexicographical ordering: $(x_1, x_2) < (y_1, y_2) \Leftrightarrow (x_1 < y_1$ or $(x_1 = y_1$ and $x_2 < y_2))$. Since cones $Q_{(0,a)}^+$ and $Q_{(0,b)}^-$ are open, their intersection, i.e. segment $a < x_2 < b$ on the $x_2$ axis, is also open. Therefore the $x_2$ axis itself and all its translates (lines $x_1 =$ const) are open sets. Hence the weakest topology of $\mathfrak{R}^2$ is the topology of a Cartesian product of $x_2$ axis (with Euclidean topology) and $x_1$ axis with *discrete* topology, $\mathfrak{R}^2 = \mathbb{R}^1 \times \mathbb{R}^1_d$. All axioms of part 1 are fulfilled.

**3.4.4.** In the case (4), Fig. 2, the ordering is not lexicographic since it is just partial ordering. But for the same reasons as in the case (3), all lines $x_1 =$ const are open and $\mathfrak{R}^2$ has a topology of a Cartesian product of $x_2$ axis (with Euclidean topology) and $x_1$ axis with *discrete* topology. All axioms of part 1 are fulfilled.

**3.4.5.** In the case (5) let us introduce oblique coordinates on the plane $\mathbb{R}^2$ directing $x, y$ axis along two edge rays of the cone $Q_0^+$. For the same reasons as in 3.4.3, each line parallel to any of the axes should be an open set. Therefore their intersection (point) is also open, and the topology in $\mathfrak{R}^2$ is discrete. It is not compatible with $LK_3$. Hence the case (5) for a linear kinematics is impossible.

**3.5.** For any finite dimension $n$ let us simulate $L = \mathfrak{R}^n$ by the points of the Euclidean space $\mathbb{R}^n$ (preserving linear structure) and consider a case when the cone $Q_0^+$ is open not only in $\mathfrak{R}^n$ but also in $\mathbb{R}^n$. An open convex cone $Q_0^+ \subset \mathbb{R}^n$ which does not contain its apex and therefore is not equal to the whole $\mathbb{R}^n$ can be represented as $Q_0^+ = C \times \mathbb{R}^k$, where $C$ is a sharp cone (i.e., it does not contain straight lines) and $0 \leq k < n$. Therefore the entire space can be represented as a Cartesian product of linear spaces $\mathfrak{R}^n = \mathfrak{R}^{n-k} \times \mathfrak{R}^k$. Closed sharp cone $C$ is open in the Euclidean space $\mathbb{R}^{n-k}$ and hence, the weakest topology introduced by $C$ in $\mathfrak{R}^{n-k}$ is Euclidean. Therefore all $(n-k)$-dimensional open balls $B$ can be taken as a base of the topological space $\mathfrak{R}^{n-k}$. In this case cylindrical tube-like neighbourhoods $B \times \mathfrak{R}^k$ of the subspace $\mathfrak{R}^k$ and all their translates will be the base of topology of the space $\mathfrak{R}^n$. Topological space $\mathfrak{R}^n$ will be represented as $\mathfrak{R}^n =$ $\mathbb{R}^{n-k} \times \mathbb{R}^{k}_{a}$, where $\mathbb{R}^{k}_{a}$ is a *k*-dimensional linear space with *antidiscrete* topology. All axioms of part 1 are fulfilled. Yet it should be noted that in this part we discuss only the weakest topology compatible with this system of axioms.

**3.6.** Let the frontier $\partial Q_0^+$ (after simulation of $\mathfrak{R}^n$ by $\mathbb{R}^n$) contain at least one ray. There is a supporting plane *P* for $\partial Q_0^+$ which contains this ray. Translating *n*-dimensional cones $Q_0^+$ and $Q_0^-$, which are open in $\mathfrak{R}^n$, one can see that both closed (in Euclidean topology) half-spaces separated by *P* are open. Therefore the hyperplane *P* itself and all its translations are open. Hence $\mathfrak{R}^n$ is represented as a Cartesian product of a line $\mathbb{R}_d$ with *discrete* topology and hyperplane *P*, the topology in *P* being determined by the cone $Q_0^+ \cap P$. Repeating with $P=\mathfrak{R}^{n-1}$ the same process and lowering dimension we shall arrive at the case 3.5, because completely discrete topology for a linear kinematics is impossible.

Hence the following theorem is proved.

**3.7. Theorem 1.** *Let in a finite-dimensional linear kinematics (L, Δ, <) the dimension of the cone $Q_0^+$ be equal to dim L, and let Δ be the weakest topology of a linear kinematics. Then L is a Cartesian product of three linear topological spaces, L=L′×A×D, L′ being a space with Euclidean topology, A with antidiscrete and D with discrete topology. If "the cone of future" $Q_0^+$ contains straight lines or a ray passing within its frontier (frontier in a sense of the Euclidean topology in L), then this decomposition is a non-trivial one, in the first case being dim A>0 and in the second case dim D>0.*

**3.8.** The discrete component in this representation is a good illustration of the concept of "parallel worlds" [4]. Let for example dim *D*>0. Then there is no path (continuous mapping, segment) connecting points from different fibers $(x_1, d_1)$ and $(x_2, d_2)$, $d_1 \neq d_2$. At the same conditions the causal relation is common for all fibers, i.e. causal link between "worlds" (fibers of space-time) is preserved.

**3.9.** If dim $Q_0^+$ <dim *L*, then in the representation *L=L′×A×D* there will be another factor *D′* (as well as *D* with discrete topology), which did not preserve causal link between its fibers.

**3.10.** R.I. Pimenov did not consider (finite dimensional) linear spaces with non-Euclidean topology, but he needed properties which in our system of axioms belong to spaces with Euclidean topology and to spaces with antidiscrete layer of maximal possible dimension (codimension 1). Appropriate causal relations are interpreted as Einstein causality (i.e. relativity theory) and Newtonian causality (preselected time axis), correspondingly. We can see that there are also other types of causality which are not equivalent to these two because they determine different topologies. Since Pimenov considered only Euclidean topology, to describe these two types of causality he had to introduce very extensive characteristics which cannot be generalized for intermediate cases.

### 4. Preordering and space of fibers of absolute simultaneity

**4.1. Definition.** *We introduce preordering $\leq$ defining it by the condition*

$$\forall a, b \in L \quad a \leq b \iff a \in \mathrm{Cl}\, Q_b^-$$

Relation $\leq$ is a relation of preordering, i.e. it is reflexive and transitive. Let us point out first the properties of the cones $Q_a^+$, $Q_a^-$.

**4.1.1.** $\forall a \in L \quad Q_a^+ = a + Q_0^+, \quad Q_a^- = a + Q_0^-$.

**4.1.2.** $\forall \alpha \in \mathbb{R}_+ \quad \alpha Q_0^+ = Q_0^+, \quad \alpha Q_0^- = Q_0^-, \quad \alpha \mathrm{Cl}\, Q_0^+ = \mathrm{Cl}\, Q_0^+, \quad \alpha \mathrm{Cl}\, Q_0^- = \mathrm{Cl}\, Q_0^-$ (self-similarity).

Therefore the relation $\leq$ is stable to addition and multiplication on scalars from $\mathbb{R}_+$.

**4.1.3.** $\forall a, b \in L \quad a \leq b \Rightarrow \forall u \in L \quad a+u \leq b+u$.

**4.1.4.** $\forall a, b \in L \quad a \leq b \Rightarrow \forall \alpha \in \mathbb{R}_+ \quad \alpha a \leq \alpha b$.

Reflexivity of $\leq$ follows from 2.3.

**4.1.5.** $\forall a \in L \quad a \leq a$.

In advance of the proof of transitivity let us point out two useful properties.

**4.1.6.** $\forall a, b, c \in L \quad a<b \ \& \ b \leq c \Rightarrow a<c$;

$\forall a, b, c \in L \quad a \leq b \ \& \ b<c \Rightarrow a<c$.

**4.1.7.** $\forall a, b, c \in L \quad a \leq b \ \& \ b \leq c \Rightarrow a \leq c$ (transitivity).

**Proof.** Let $a \in \text{Cl } Q_b^-$, $b \in \text{Cl } Q_c^-$, $U$ be any neighbourhood of a point $a$. Then there is $u \in U \cap Q_b^- \neq \emptyset$, and due to (6) $u<b \leq c \Rightarrow u<c \Rightarrow u \in Q_c^- \Rightarrow U \cap Q_b^- \subset Q_c^- \Rightarrow U \cap Q_c^- \neq \emptyset$, i.e. any neighbourhood of a point $a$ intersects with $Q_c^-$. Therefore $a \in \text{Cl } Q_c^-$, i.e. $a \leq c$.

**4.1.8.** For any $a \in L$ the sets $\text{Cl } Q_a^+$, $\text{Cl } Q_a^-$ are convex cones with apex $a$ (apex belongs to the cone due to (5)). (Proof is quite analogous to the proofs of 2.6 and 2.7 with substitution of $<$ with $\leq$).

**4.1.9.** $\forall a \in L \quad \text{IntCl } Q_a^+ = Q_a^+$, $\text{IntCl } Q_a^- = Q_a^-$.

**Proof.** Let us check as an example the first statement. It is obvious that $\text{IntCl } Q_a^+ \supset Q_a^+$. Prove that no point $u \in \text{Cl } Q_a^+$, $u \notin Q_a^+$ can belong to $\text{IntCl } Q_a^+$. Any point $u$ with these properties should belong to $\partial Q_a^+$. Due to LK$_3$ in each neighbourhood $U$ of the point $u$ there is $v<u$. If $v \in \text{Cl } Q_a^+$, then $a \leq v<u$, and that makes a contradiction with $u \notin Q_a^+$. Therefore in any neighbourhood of the point $u$ there are points which do not belong to $\text{Cl } Q_a^+$, i.e. $u \notin \text{IntCl } Q_a^+$.

**4.2. Definition.** *The preordering $\leq$ defines the relation of equivalence $\sim$ on $L$*:

$\forall a, b \in L \quad a \sim b \Leftrightarrow a \leq b \ \& \ b \leq a$.

In physics this relation is called *absolute simultaneity*. It is symmetric by definition, and reflexive and transitive due to 4.1.5 and 4.1.7.

**4.3. Definition.** *Class of points equivalent to a given point $c$ is designated as $\tilde{c} := \{b \in L \mid c \sim b\}$. This set can be also called fiber of simultaneity of the point $c$.*

We would like to point out connection of this term with the results of part 3. Let $L$ be a Cartesian product of the three linear topological spaces, $L=L' \times A \times D$ with Euclidean, antidiscrete and discrete topology, respectively. Then the class of points equivalent to $a=(a_{L'}, a_A, a_D)$ is $\tilde{a} = a_{L'} \times A \times a_D$, i.e. if the cone $Q_0^+$ contains straight lines then the fiber of simultaneity of the point $a$ consists exactly of the sum of all translates of these lines which pass through the point $a$. Indeed, since in definition 4.1 there is a closure operation, the entire set $A$ (or more exactly, its translate) is contained in $\tilde{a}$. On the other hand, the fiber of simultaneity cannot be bigger because the cone $Q_0^+ \cap L'$ is sharp.

**4.3.1.** For any $a \in L$ the sets $Q_a^+$, $Q_a^-$, $\text{Cl } Q_a^+$, $\text{Cl } Q_a^-$ consists of complete classes of equivalence.

**4.3.2.** Class $\tilde{0}$ is a linear subspace in $L$. (Under conditions of Theorem 1 $\tilde{0}$ is identical with $A$).

**4.3.3.** For any $a \in L \quad \tilde{a} = a + \tilde{0}$.

**4.3.4.** For any $a \in L$ the set $\tilde{a}$ is closed ($\tilde{a} = \text{Cl } Q_a^+ \cap \text{Cl } Q_a^-$).

**4.3.5.** For any $a \in L$  Int $\tilde{a} = \emptyset$. (It follows from 4.3.4 and 4.1.9). (Under conditions of Theorem 1 it means that dim $L' \geq 1$).

### 4.4. Space of fibers of absolute simultaneity.

**Definition.** *The set of classes of equivalence ~ we designate as $\tilde{L} := L/\sim$.*

Since all classes of equivalence are the translates of the subspace $\tilde{0}$ (4.3.2, 4.3.3), the set $\tilde{L}$ inherits from $L$ its linear structure, $L$ being a Cartesian product $L \approx \tilde{L} \times \tilde{0}$.

$\tilde{L}$ as factor-space inherits from $L$ topological structure ($\Delta$). Let us designate $\pi: L \to \tilde{L}$ - factorization mapping (projection). Then $A \subset \tilde{L}$ is open in $\tilde{L}$ iff $\pi^{-1}(A)$ is open in $L$.

Due to Theorem 1 of part 3.7 the following statement is obvious: $\tilde{L}$ *is a Hausdorff space.* Indeed factorspace is Hausdorff even in the weakest topology of a finite dimensional linear kinematics. However this statement is true also for infinite dimensional linear kinematics. The proof is relatively simple and we shall not discuss it here.

**4.4.1. Lemma.** $\pi: L \to \tilde{L}$ *is an open mapping, i.e. if $A \subset L$ is open, then $\pi(A) \subset \tilde{L}$ is also open.*

**Proof.** We must add to $A$ all classes of equivalence for all elements from $A$ and prove that thus extended set $B := \pi^{-1}(\pi(A))$ is open.

Let $b \in B$, then $\exists a \in A$ so that $a \sim b$ and there is an open set $U$ so that $a \in U \subset A$. Consider an open (due to axiom $H_1$) set $V := (b-a)+U$, $b \in V$. We should prove that $V \subset B$. Indeed for any $v \in V$ $\exists u \in U$ $v = b-a+u$. Since $a \sim b \Rightarrow a-b \sim 0 \Rightarrow v \sim u$ and $u \in A$, then $v \in B$. Hence, any point $b \in B$ belong to $B$ together with its neighbourhood, i.e. $B$ is open.

**4.4.2.** We had already noted that equivalence ~ is stable to addition of vectors and multiplication by scalars. The ordering < is also stable to this equivalence. It immediately follows from 4.1.6.

1. $\forall a,b,c \in L$  $a<b$ & $b \sim c$ $\Rightarrow$ $a<c$.
2. $\forall a,b,c \in L$  $a \sim b$ & $b<c$ $\Rightarrow$ $a<c$.

Therefore factorspace $\tilde{L}$ inherits ordering from $L$ (<).

**Definition.** *Let $\tilde{a}, \tilde{b} \in \tilde{L}$, then $\tilde{a} \stackrel{\sim}{<} \tilde{b} \Leftrightarrow a<b$.*

**4.4.3. Theorem 2.** *Linear space $\tilde{L}$ with ordering $\stackrel{\sim}{<}$ and factortopology $\tilde{\Delta}$ is a linear kinematics.*

**Proof.** We must verify that all axioms of a linear kinematics are fulfilled.
The continuity of addition follows from the commutativity of the diagram

$$
\begin{array}{ccc}
L \times L & \xrightarrow{+} & L \\
\downarrow (\pi,\pi) & & \downarrow \pi \\
\tilde{L} \times \tilde{L} & \xrightarrow{+} & \tilde{L}
\end{array}
\qquad \forall\, a, b \in L \quad \pi(a) + \pi(b) = \pi(a+b)
$$

Let $A \subset \tilde{L}$ be open. Then $\pi^{-1}(A) \subset L$ is also open. Addition in $L$ is continuous, therefore the pre-image $B$ of a set $\pi^{-1}(A)$ at the mapping $+: L \times L \to L$ is open in $L \times L$. $\pi: L \to \tilde{L}$ is an open mapping, therefore $(\pi,\pi): L \times L \to \tilde{L} \times \tilde{L}$ is also an open mapping. Then image $C := (\pi,\pi)(B)$ of an open set $B$ is open in $\tilde{L} \times \tilde{L}$. Since the above diagram is commutative, the set $C$ is identical with the pre-image of $A$ at mapping $+: \tilde{L} \times \tilde{L} \to \tilde{L}$. Hence this pre-image is open and addition in factorspace $\tilde{L}$ is continuous.

Let us check the axiom $H_2$: for any $\alpha \in \mathbb{R}$ the mapping $\alpha.: \tilde{L} \to \tilde{L}: \tilde{a} \mapsto \alpha \tilde{a}$ is continuous. If $\alpha=0$, then this is a constant mapping and therefore continuous. Let $\alpha \neq 0$. Corresponding diagram is commutative, since $\pi \circ \alpha. = \alpha. \circ \pi$. The same reasons as above prove our statement.

Proofs of other axioms are trivial.

**4.4.4.** It should be noted that fibers of absolute simultaneity appear in linear kinematics due to very different physical reasons. Thus in Galilean kinematics $\mathbb{R}^3 \times \{t\}$, where the cone of future is the half-space $t>0$, the fiber of absolute simultaneity is $\mathbb{R}^3$. In kinematics $\mathbb{R}^4 \times \mathbb{R}$ (Minkovskii space with the fiber $\mathbb{R}$ over each point used in [5, p. 399] to simulate electromagnetism) the fiber of simultaneity is $\mathbb{R}$.

## 5. Metric linear kinematics

**5.1. Definitions.** In physics definition of ordering using some function (for example, which is positive iff its argument is 'greater' than zero point) would be more convenient than an abstract definition used in part 1. Since ordering must be transitive, this function should satisfy time inequality (triangle inequality with opposite sign). We shall build a linear kinematics where both ordering and topology are defined by such function. Most of its properties will prove to be determined by the domain of this function.

**Definition.** *Let $L$ be a linear space over the field $\mathbb{R}$, $Q_0^+$ – convex cone in $L$ with apex $0$, $0 \notin Q_0^+$. Designate $Q_0^- := (-1) Q_0^+$, $F := Q_0^+ \cup Q_0^- \cup \{0\}$. We shall consider only non-empty cones with the following property:*

**M$_1$.** $\forall a \in Q_0^+ \;\; \forall b \in L \;\; \exists \alpha \in \mathbb{R}_+ \;\; a + \alpha b \in Q_0^+$

*Function $f: F \to \mathbb{R}$ which is strictly positive on $Q_0^+$ is called a time-like norm (or a pseudonorm) iff it satisfies the following conditions:*

**M$_2$.** $\forall a, b \in F \;\; a>0 \;\&\; b>0 \Rightarrow f(a+b) \geq f(a)+f(b)$ *(Einstein inequality, or time inequality).*

**M$_3$.** $\forall a \in F \;\; \forall \alpha \in \mathbb{R} \;\; f(\alpha a) = \alpha f(a)$ *(uniformness).*

Since $Q_0^+$ is a convex cone, $M_2$ is correct ($a \in Q_0^+ \;\&\; b \in Q_0^+ \Rightarrow a+b \in Q_0^+$). $M_3$ is correct because $F$ is a cone symmetrical with respect to the zero point. Besides $f(0) = f(0x) = 0 f(x) = 0$ since the cone $Q_0^+$ is not empty.

Each convex cone defines relation of partial ordering on $L$: for any $a, b \in L$

$$a < b \Leftrightarrow b - a \in Q_0^+.$$

The relation $<$ is transitive ($\forall a, b, c \in L \;\; a<b \;\&\; b<c \Rightarrow b-a \in Q_0^+ \;\&\; c-b \in Q_0^+ \Rightarrow c-a \in Q_0^+ \Rightarrow a<c$) and antireflexive ($0 \notin Q_0^+ \Rightarrow \forall a \in L \;\; \neg(a<a)$). Thus the axioms $O_1$ and $O_2$ of part 1 are fulfilled.

Because $f$ is strictly positive only on $Q_0^+$, one can give another definition of partial ordering which is equivalent to the first:

$$\forall a, b \in L \;\; a<b \Leftrightarrow b-a \in F \;\&\; f(b-a)>0.$$

Under this definition transitivity of ordering $<$ follows from Einstein inequality, antireflexivity – from the fact that $f(0)=0$. Axiom $M_1$ can be formulated in a pure function form:

**M$'_1$.** $\forall a \in F \;\; f(a)>0 \Rightarrow \forall b \in L \;\; \exists \alpha \in \mathbb{R}_+ \;\; a+\alpha b \in F \;\&\; f(a+\alpha b)>0$.

**Definition.** *Designate $W := \{(a, b) \mid a, b \in L, b-a \in F\} \subset L \times L$. We shall call function $\rho: W \to \mathbb{R}$, $\rho(x, y) := f(x-y)$ the timelike metric (or pseudometric) in $W$.* (In physics pseudometric has a sense of oriented distance between two events separated by the timelike interval).

Einstein inequality for the timelike metric has the form

**M$_2$.** $\forall a, b, c \in L \ \ a>b \ \& \ b>c \ \Rightarrow \ \rho(a, c) \geq \rho(a, b)+\rho(b, c)$

Busemann ([3], pp. 5,7) wrote that pseudometric which satisfies Einstein inequality did not define a topology on a space $L$. Pimenov ([5], p.257) also assumed it. We shall prove that with the additional axioms $M_1$ and $M_3$ pseudometric defines a topology on space $L$, and this topology is quite different from the Euclidean one. We also shall find an additional condition which is sufficient and necessary for a linear space $L$ with pseudonorm $f$ to be a linear kinematics.

### 5.2. Corollaries.

**5.2.1.** $\forall a \in Q_0^+ \ \ \forall b \in L \ \ \forall \alpha \in \mathbb{R}_+ \ \ a+\alpha b \in Q_0^+ \ \Rightarrow \ (\forall \beta \in \mathbb{R}_+ \ \ \beta < \alpha \Rightarrow a+\beta b \in Q_0^+)$.

It immediately follows from the convexity of the cone $Q_0^+$. This corollary can be formulated also in a function form. *Let $a \in F$, $b \in L$, $f(a)>0$, $\alpha \in \mathbb{R}_+$, $a+\alpha b \in F$, $f(a+\alpha b)>0$. Then $\forall \beta \in \mathbb{R}_+ \ \ \beta < \alpha \Rightarrow a+\beta b \in F \ \& \ f(a+\beta b)>0$.*

**5.2.2.** $\forall a \in F \ \ \forall \varepsilon \in \mathbb{R}_+ \ \ f(a) > \varepsilon \ \Rightarrow \ \forall b \in L \ \ \exists \alpha \in \mathbb{R}_+ \ \ a+\alpha b \in F \ \& \ f(a+\alpha b)>\varepsilon$.

**Proof.** Let us take $\delta>0$ and represent $a$ as $a=a-\delta a+\delta a$. Due to $M_1$ there is $\beta \in \mathbb{R}_+$ so that $f(\delta a+\beta b)>0$. Therefore, $f(a+\beta b) \geq f(a-\delta a) + f(\delta a+\beta b) > (1-\delta)f(a) > (1-\delta)\varepsilon$. The left part of this inequality does not depend on $\delta$. If we take $\sup_{\delta \in \mathbb{R}_+}$ we get $f(a+\beta b) \geq \varepsilon$.

Then for any $\alpha \in \mathbb{R}_+ \ \ \alpha < \beta \ \Rightarrow \ f(a+\alpha b) = f\left(\frac{\alpha}{\beta}\left(\frac{\beta}{\alpha}a+\beta b\right)\right) = \frac{\alpha}{\beta} f\left(a+\beta b+\left(\frac{\beta}{\alpha}-1\right)a\right) \geq$

$\geq \frac{\alpha}{\beta}\left(f(a+\beta b)+\left(\frac{\beta}{\alpha}-1\right)f(a)\right) \geq \frac{\alpha}{\beta}\varepsilon+\left(1-\frac{\alpha}{\beta}\right)f(a) > \varepsilon$, since at $\alpha=\beta$ the latter statement is an equality, and in the left part of it there is a linear function of $\alpha$, which increases when $\alpha$ decreases.

**5.2.3.** *Let $a \in F$, $b \in L$, $\varepsilon \in \mathbb{R}_+$, $f(a)>\varepsilon$, $\alpha \in \mathbb{R}_+$, $a+\alpha b \in F$, $f(a+\alpha b)>\varepsilon$. Then for any $\beta \in \mathbb{R}_+ \ \ \beta < \alpha \ \Rightarrow \ a+\beta b \in F \ \& \ f(a+\beta b)>\varepsilon$.* (This was actually proved in the proof of the previous statement).

### 5.3. Definition of topology by means of pseudometric.

**5.3.1. Definition.** *The set $B_\varepsilon(a):=\{x \in L \ | \ x-a \in F \ \& \ \rho(x, a)>\varepsilon\}$ will be called extraball with radius $\varepsilon$ and center $a$.*

It is more natural to search a topology with the base of extraballs which are convex sets (obvious) than to consider balls which are not convex for a space with pseudometrics.

**Theorem 3.** *The set of extraballs $\Gamma:=\{B_\varepsilon(a) \ | \ a \in L, \ \varepsilon \in \mathbb{R}_+\}$ is a base of a topology on $L$.*

**Proof.** 1. Extraballs from $\Gamma$ cover all $L$. Let us take an arbitrary point $x \in L$ and vector $u \in Q_0^-$. It is obvious that $x+u<x$ and if we designate $f(u):=-2\varepsilon$, $\varepsilon \in \mathbb{R}_+$, then $x \in B_\varepsilon(x+u)$.

2. Intersection of two sets from $\Gamma$ can be represented as a sum of sets from $\Gamma$. Let $c \in B_\varepsilon(a) \cap B_\delta(b)$. We must check that $\exists h \in L \ \exists \alpha \in \mathbb{R}_+$ so that $c \in B_\alpha(h) \subset B_\varepsilon(a) \cap B_\delta(b)$. Let us take $u \in Q_0^-$. Due to 5.2.2 there is $r_1 \in \mathbb{R}_+$ and $r_2 \in \mathbb{R}_+$ so that $f(c+r_1u-a)>\varepsilon$ and $f(c+r_2u-b)>\delta$. Assigning $r:=\min\{r_1, r_2\}$, $h:=c+ru$, $f(h-c):=-2\alpha$, $\alpha \in \mathbb{R}_+$, we have that $h \in B_\varepsilon(a) \cap B_\delta(b)$, $h<c$ and $c \in B_\alpha(h)$. If $x \in B_\alpha(h)$, then $f(x-a) \geq f(x-h)+f(h-a)>\varepsilon$ and $f(x-b) \geq f(x-h)+f(h-b)>\delta$, therefore $B_\alpha(h) \subset B_\varepsilon(a) \cap B_\delta(b)$.

**Definition.** *The topology on the space $L$ with the base $\Gamma$ is called the timelike topology determined by the pseudonorm $f$.*

**5.3.2.** It is well known that in a finite dimensional linear space all norms determine the same topology. We shall prove (without using the condition of finite dimension) that *the timelike topology determined by the pseudonorm f actually depends only on the domain of f*. Let us designate $Q_a^+ := a + Q_0^+$, $\Gamma' := \{ Q_a^+ \mid a \in L \}$.

**Lemma 1.** *The set of cones $\Gamma'$ is a base of a topology on L.*
Proof is quite analogous to the proof of Theorem 3.

**Lemma 2.** *In topology with the base $\Gamma$ all cones from $\Gamma'$ are open.*
**Proof.** This statement immediately follows from the equality $Q_0^+ = \bigcup_{\varepsilon \in \mathbb{R}_+} B_\varepsilon(0)$ and the fact that the mapping of addition is continuous under this set of axioms ($M_1$–$M_3$) (see the proof of Theorem 5 in 5.4.2).

**Lemma 3.** *In topology with the base $\Gamma'$ all extraballs from $\Gamma$ are open.*
**Proof.** We must check that for any $x \in B_\varepsilon(a)$ there is $y \in L$ so that $x \in Q_y^+ \subset B_\varepsilon(a)$. It is sufficient to take $u \in Q_0^-$ and choose (using 5.2.2) $\alpha \in \mathbb{R}_+$ so that $x + \alpha u \in B_\varepsilon(a)$. It is obvious that $y = x + \alpha u$.

Hence the following theorem was proved.

**Theorem 4.** *Topologies with the bases $\Gamma$ and $\Gamma'$ are the same.*

**5.3.3.** This topology is quite different from the standard topology of a metric space. For example, it cannot be Hausdorff. If $a < b$ and $a \in B_r(x)$, then $b \in B_r(x)$. ($a < b$ & $a \in B_r(x)$ $\Rightarrow$ $f(b-x) \geq f(b-a) + f(a-x) > r$ $\Rightarrow$ $b \in B_r(x)$). Also, multiplication by $-1$ in this topology is not continuous. This is related with the fact that all sets from $\Gamma$ are unlimited towards the direction of "time increasing" $Q_0^+$ and limited on the other side.

**5.4. Strengthening of topology and construction of a linear kinematics.** To build a linear kinematics the increasing of the timelike topology is necessary.

**5.4.1. Lemma.** *The set $\Delta := \{U \cap (-V) \mid U, V \in \Gamma\}$ is a base of a topology on L, this topology being stronger than the timelike topology.*

Proof of this lemma is a quite direct one (using the fact that $\Gamma$ is a base of topology on $L$).

**Example.** Let us take $\mathbb{R}$ as a linear space and $f = \text{id}$ as a pseudonorm. Then the timelike topology on $\mathbb{R}$ is a topology with the base of rays $\{(x, +\infty) \mid x \in \mathbb{R}\}$, and the topology with the base $\Delta$ is Euclidean.

**Definition.** *The topology on L with the base $\Delta$ is called the topology determined by the pseudonorm f.*

**5.4.2. Theorem 5.** *Linear space L with the topology determined by the pseudonorm f and with the ordering determined by the same pseudonorm is a linear kinematics.*

**Proof.** Axioms $O_1$, $O_2$ have been already checked in definition of partial ordering on $L$ (5.1). Let us check the axiom $H_1$. We designate the sets from the base $\Delta$ as $B_{r,q}(a,b) := B_r(a) \cap -B_q(b)$, where $a, b \in L$, $r, q \in \mathbb{R}_+$.

**Lemma.** $\forall a,b,c,d \in L \ \forall r,q,p,t \in \mathbb{R}_+ \ B_{r,q}(a,b) + B_{p,t}(c,d) \subset B_{r+p,q+t}(a+c,b+d)$.
Proof is quite direct and uses only the time inequality.

Continue the proof of the theorem 5. Let a set $U \subset L$ be open. We must verify that its pre-image at mapping $+: L \times L \to L$ is open in $L \times L$, i.e. for any $(x, y) \in L \times L$ $x+y \in U$ $\Rightarrow$ $\exists V, W \subset L$ – open sets so that $x \in V$, $y \in W$, $V + W \subset U$. One can assume that $U, V, W \in \Delta$. Let $U = B_{\alpha,\beta}(u, v)$, $e \in Q_0^+$ and $\gamma \in \mathbb{R}_+$. Considering $\rho(x, x-\gamma e) = f(\gamma e) = \gamma f(e) > 0$ let us select $\gamma_1 \in \mathbb{R}_+$ so that $f(\gamma_1 e) < \alpha$ and let $\gamma < \gamma_1$. Then $x \in B_r(x-\gamma e)$, where $f(\gamma_1 e) =: 2r < 2\alpha$. Consider $\rho(y, u-(x-\gamma e)) = f(x+y-u-\gamma e)$, where according to conditions $f(x+y-u) > \alpha$. Due to $M_1$ there is $\gamma_2 \in \mathbb{R}_+$ so that $f(x+y-u-\gamma_2 e) > \alpha$. Let us select $\gamma = \min\{\gamma_1, \gamma_2\}$ and assign $a := x - \gamma e$. Then $x \in B_r(a)$ and $y \in B_{\alpha-r}(c)$, where $c := u - a$.

Analogous, there are $q<\beta$ and $b \in L$ so that $x \in -B_q(b)$ and $y \in -B_{\beta-q}(d)$, where $d:=v-b$. Now the inclusion $B_{r,q}(a,b)+B_{\alpha-r, \beta-q}(c,d) \subset B_{\alpha,\beta}(u,v)$ is ensured by Lemma.

Let us check the axiom $H_2$. Since multiplication on a scalar $\alpha \in \mathbb{R}$ with $\alpha \neq 0$ is a one-to-one mapping, the continuity of multiplication is equivalent to the following proposition: *if $U \subset L$ is open, then for any $\alpha \in \mathbb{R}$ ($\alpha \neq 0$) $\alpha U \subset L$ is also open*. It is enough to check this statement for the sets from $\Delta$ only. Let $U:=B_r(a) \cap -B_q(b)$. If $\alpha \in \mathbb{R}_+$, then $\alpha B_r(a) = B_{\alpha r}(\alpha a)$. If $\alpha \in \mathbb{R}_-$, then $\alpha B_r(a) = -B_{|\alpha|r}(|\alpha|a)$. These sets are open.

Stability of ordering < to addition (axiom $LK_1$) follows directly from definition. Stability of ordering to multiplication ($LK_2$) follows from the axiom $M_3$ and definition of topology determined by the pseudonorm $f$.

Let us check the axiom $LK_3$: in any neighbourhood of any point there is a next point. It is sufficient to check this statement for the sets from $\Delta$. Let $a \in B_{r,q}(u, v)$ and $b \in Q_0^+$. Then for any $\alpha \in \mathbb{R}_+$ $a+\alpha b > a$. Statement 5.2.2 implies that $\exists \alpha_1 \in \mathbb{R}_+$ $\rho(-a-\alpha_1 b, v) > q$. Then $a+\alpha_1 b \in B_{r,q}(u, v)$. This is the "next point" belonging to the same neighbourhood.

Let us check the axiom $LK_4$. We know (see the proof of Lemma 2 in 5.3.2) that $Q_0^+$ is open. By definition of the base $\Delta$, multiplication on $-1$ is continuous, therefore $Q_0^- = -Q_0^+$ is also open.

**5.4.3.** Thus we proved that for a space $L$ with pseudonorm $f$ to be a linear kinematics it is necessary and sufficient that the mapping of multiplication by $-1$ be continuous. In physics it corresponds to the requirement that the reversion of time is an allowed operation.

**Corollary.** *Let $L$ be a linear space, $f: F \to \mathbb{R}$ — pseudonorm in $F$, $F:= Q_0^+ \cup Q_0^- \cup \{0\}$. Then the topology in $L$ determined by the pseudonorm $f$ is identical with the weakest topology of the linear kinematics $L$ with the cone of future $Q_0^+$.* (It immediately follows from 5.4.2 and 5.3.2).

**5.5. The continuity of the pseudonorm.**

**Theorem 6.** *Function $f$ is continuous.*

**Proof.** It is sufficient to consider only the pre-images of rays $(\varepsilon, +\infty)$ and $(-\infty, \varepsilon)$ (where $\varepsilon \in \mathbb{R}$) and prove that they are open in $F$. If $\varepsilon \in \mathbb{R}_+$, then $f^{-1}((\varepsilon, +\infty)) = B_\varepsilon(0)$. If $\varepsilon \in \mathbb{R}_-$, then $f^{-1}((-\infty, \varepsilon)) = -B_{|\varepsilon|}(0)$. If the ray contains zero point, then $f^{-1}((-\infty, \varepsilon)) \cap F = F \setminus \mathrm{Cl}\, B_\varepsilon(0)$ for $\varepsilon \in \mathbb{R}_+$ and $f^{-1}((\varepsilon, +\infty)) \cap F = F \setminus \mathrm{Cl} -B_{|\varepsilon|}(0)$ for $\varepsilon \in \mathbb{R}_-$. Therefore these pre-images are open in $F$.

**5.6. The space of fibers of absolute simultaniety.**

**Definition.** *Linear kinematics $(L, \Delta, <)$ is called metric linear kinematics, iff there is a pseudonorm $f: F \to \mathbb{R}$ so that the topology $\Delta$ is determined by $f$ and ordering < is compatible with $f$ (i.e. $\{x \in L \mid x > 0\} = \{x \in L \mid f(x) > 0\}$).*

**5.6.1. Theorem 7.** *Let $(L, f)$ be a metric linear kinematics. Then factorspace $\tilde{L}$ is also a metric linear kinematics, and there is a topological immersion $\tilde{L} \to L$. For any topological immersion $\phi: \tilde{L} \to L$ there is a homeomorphism $L \cong \phi(\tilde{L}) \times \tilde{0}$, where $\tilde{0} \subset L$ has antidiscrete and $\phi(\tilde{L})$ has a Hausdorff topology.*

**Proof.** Let $(L, f)$ be a metric linear kinematics, $\tilde{L}$ — its factorkinematics. Prove that $\tilde{L}$ is also a metric kinematics with a pseudonorm defined by the following rule. Let us designate factorization mapping as $\pi: L \to \tilde{L}$, $\tilde{F}:= \pi(F)$. For any $\tilde{a} \in \tilde{F}$ we define $\tilde{f}(\tilde{a}):=f(a)$.

This definition is correct: if $a \sim b$, then $f(a)=f(b)$. Let on the contrary $a \sim b$ and $f(a) \neq f(b)$. One can assume that $f(a)>f(b)>0$ ($f(a)$ and $f(b)$ cannot have different signs because ordering $<$ is stable to equivalence $\sim$). There is an $\varepsilon \in \mathbb{R}_+$ so that $f(a)>\varepsilon>f(b)$. Therefore there is $\alpha \in \mathbb{R}_+$ so that $f(\alpha a)=\varepsilon$. Hence $f(a)>f(\alpha a) \Rightarrow (1-\alpha)f(a)>0 \Rightarrow f(a-\alpha a)>0$. Yet $a \sim b \Rightarrow a-\alpha a \sim b-\alpha a \Rightarrow f(b-\alpha a)>0$. Therefore $f(b)=f(b-\alpha a+\alpha a) \geq f(b-\alpha a)+f(\alpha a)>\varepsilon$ – contradiction with assumption $f(b)<\varepsilon$.

Let us continue the proof of the theorem 7. We must check that $\tilde{f}$ is a pseudonorm. Designate $\tilde{Q}_0^+ := \pi(Q_0^+)$, $\tilde{Q}_0^- := \pi(Q_0^-)$. Projection $\pi$ is a linear mapping (see 4.3.2 and 4.3.3), therefore $\tilde{Q}_0^+$ is a convex cone in $L$ with apex $\tilde{0}$, $\tilde{0} \notin \tilde{Q}_0^+$.

Let us check M$_1$. Let $\tilde{a} \in \tilde{Q}_0^+$, $\tilde{b} \in \tilde{L}$. Then for any representatives in these classes $\exists \alpha \in \mathbb{R}_+$ $a+\alpha b \in Q_0^+$. Projecting it into $\tilde{L}$ we have $\tilde{a}+\alpha \tilde{b} \in \tilde{Q}_0^+$.

Let us check M$_2$. If $a>0$ and $b>0$, then for any representatives in these classes $a>0$, $b>0$ and $f(a+b) \geq f(a)+f(b)$. Since $f = \tilde{f} \circ \pi$ and $\pi$ is a linear mapping, we have time inequality in factorspace. Axiom M$_3$ can be verified in the same way. Hence the function $\tilde{f}$ is a pseudonorm.

It is obvious that $\tilde{f}$ is strictly positive only on $\tilde{Q}_0^+$, therefore ordering $\tilde{<}$ is compatible with $\tilde{f}$.

Pseudonorm $\tilde{f}$ defines a topology on the factorspace $\tilde{L}$ which is identical with the factortopology $L/\sim$, because bases of the two topologies match: $\tilde{B}_\varepsilon(\tilde{a}) \cap -\tilde{B}_\delta(\tilde{b}) = \pi(B_\varepsilon(a) \cap -B_\delta(b))$. Hence ($\tilde{L}$, $\tilde{\Delta}$, $\tilde{<}$) is a metric linear kinematics.

To construct an immersion $\phi$: $\tilde{L} \to L$ we must fix some choice of representatives $\tilde{a} \mapsto a=:\phi(\tilde{a})$. It is an injection. Continuity of $\phi$ follows from the equality $\phi^{-1}(B_\varepsilon(a) \cap -B_\delta(b)) = \tilde{B}_\varepsilon(\tilde{a}) \cap -\tilde{B}_\delta(\tilde{b})$. Let us consider the abbreviation of $\phi$, ab $\phi$: $\tilde{L} \to \text{im } \phi$. The inverse mapping (ab $\phi)^{-1}$: im $\phi \to \tilde{L}$ is continuous, because (ab $\phi)^{-1} = \pi$ |im $\phi$. Therefore $\phi$ is an immersion.

Now let $\phi$: $\tilde{L} \to L$ be an arbitrary topological immersion. Then homeomorphism $\psi$: $L \to \phi(\tilde{L}) \times \tilde{0}$ can be defined as $\psi(a):=(\phi(\tilde{a}), a-\phi(\tilde{a}))$, $a \in L$. The base of topology in $L$ is identical with the base of topology in $\phi(\tilde{L}) \times \tilde{0}$. Also there is a homeomorphism $L \to \tilde{L} \times \tilde{0}$. It can be defined by $\tilde{a} \mapsto (\tilde{a}, a-\phi(\tilde{a}))$.

Due to 4.3.1 all open (and closed) sets of the topology of a metric kinematics consist of complete classes of equivalence. In particular on the subset $\tilde{0} \subset L$ the topology is antidiscrete. Indeed this set is closed (4.3.4) and consists of the only class of equivalence. Therefore it does not contain non-trivial closed subsets. Theorem 7 has been proved.

**5.6.2. Corollary.** *Let (L, f) be a finite dimensional metric linear kinematics. Then L is a Cartesian product of two linear spaces, $L=E \times A$, where A has antidiscrete and E Euclidean topology.*

**Proof.** Due to Theorem 7 and 4.4 $L=H \times A$, where $H$, $A$ are linear spaces with Hausdorff ($H$) and antidiscrete ($A=0$) topology. Due to Theorem 1 and 5.4.3 $L=E \times D \times A$, where $E$, $D$, $A$ are linear spaces with Euclidean, discrete and antidiscrete topology, respectively. Therefore $H=E \times D$. But due to M$_1$ discrete space cannot be presented in this decomposition, i.e. dim $D=0$. Let on the contrary dim $D>0$. Then either there is a set from the base $\Delta$ so that aff($B_\varepsilon(a) \cap -B_\delta(b)) \neq L$, or the cone $Q_0^+$ contains rays belonging to $\partial Q_0^+$ (where frontier is taken in Euclidean topology of the linear space $L$). In the second case contradiction with $M_1$ is obvious. In the first case there are points $x \in B_{\varepsilon,\delta}(a, b)$, $y \in L \backslash \text{aff } B_{\varepsilon,\delta}(a, b)$. Due to 5.2.2 and 5.2.3, $\exists \gamma \in \mathbb{R}_+$ $x + \gamma(y-x) \in B_{\varepsilon,\delta}(a, b)$, but it is impossible because $y \notin \text{aff } B_{\varepsilon,\delta}(a, b)$.

Thus factortopology of a finite dimensional metric linear kinematics is always Euclidean. Euclidean spaces with partial ordering were also studied by A.D. Alexandrov [1, 2].

**5.6.3.** The case of "parallel worlds" can be partially described in terms of metric linear kinematics if the condition of axiom $M_1$ is changed:

**M**$_4$. $\forall a \in Q_0^+$ $\forall b \in F$ $\exists \alpha \in \mathbb{R}_+$ $a + \alpha b \in Q_0^+$, or in a functional form

**M'**$_4$. $\forall a \in F$ $f(a) > 0$ $\Rightarrow$ $\forall b \in F$ $\exists \alpha \in \mathbb{R}_+$ $a + \alpha b \in F$ & $f(a + \alpha b) > 0$.

In this statement condition "$\forall b \in L$" is replaced by the weaker one "$\forall b \in F$". Corollaries from 5.2 can be easily reformulated considering this change (replacing "$b \in L$" with "$b \in F$"). Corollary from 5.6.2 will be changed:

**Proposition.** *Let $(L, f)$ be a finite dimensional metric linear kinematics, satisfying axioms $M_4$, $M_2$, $M_3$. Then $L$ is a Cartesian product of three linear spaces, $L = E \times D \times A$ with Euclidean, discrete and antidiscrete topology, respectively.*

It should be noted that layers $E \times A$ are not linked by the relation of ordering, as it is in 3.9.

**5.6.4. Example.** The most well-known example of a metric linear kinematics is a Minkovskii space. Let us consider the space $L = \mathbb{R}^4$ and the function $f(x, y, z, t) = \text{sgn}(t)\sqrt{t^2 - x^2 - y^2 - z^2}$ ($\text{sgn}(0) := 0$). Designate $F := \text{Supp} f \cup \{0\}$. It is a cone and function $f$ is a timelike norm (the cone $Q_0^+$ is a set where $f$ is positive, axioms $M_{1-3}$ are fulfilled). Function $f$ determines in $L$ Euclidean topology.

**5.7. Question.** It is obvious that under the conditions of axioms $M_1$-$M_3$ the function $f$ is concave on both $Q_0^+$ and $Q_0^-$. Therefore the following question arises: *let $f$ be a strictly positive concave function on $Q_0^+$ defined on a cone $F = Q_0^+ \cup Q_0^- \cup \{0\}$ where $Q_0^+$ satisfies the axiom $M_1$. Is it true that $f$ determines a base of a topology in a linear space $L$?*

### Acknowledgement
The author is grateful to V.A. Zalgaller for useful discussions.